\documentstyle[aps,epsf,epsfig]{revtex}

\newcommand{\bq}{\begin{equation}}
\newcommand{\eq}{\end{equation}}
\newcommand{\bqn}{\begin{eqnarray}}
\newcommand{\eqn}{\end{eqnarray}}
\newcommand{\nb}{\nonumber}
\newcommand{\lb}{\label}
\newcommand{\noi}{\noindent}
\begin{document}
\title{ Gravitational Collapse of Massless Scalar Field with Negative
Cosmological Constant in (2+1) Dimensions}
\author{R. Chan\thanks{E-mail: chan@on.br} $^1$,
M. F. A. da Silva\thanks{E-mail: mfas@dtf.if.uerj.br} $^2$ ,
Jaime F. Villas da Rocha\thanks{E-mail: roch@dft.if.uerj.br} $^2$}
\address{$^1$ Coordenadoria de Astronomia e Astrof\'{\i}sica,
Observat\'orio Nacional, Rua General Jos\'e Cristino 77, S\~ao
Crist\'ov\~ao, CEP 20921--400, Rio de Janeiro, RJ, Brazil \\
$^2$ Departamento de F\' {\i}sica Te\' orica,
Universidade do Estado do Rio de Janeiro,
Rua S\~ ao Francisco Xavier $524$, Maracan\~ a,
CEP 20550--013, Rio de Janeiro, RJ, Brazil} 

\date{\today}

\maketitle

\begin{abstract}
The $2+1$-dimensional geodesic circularly symmetric solutions of
Einstein-massless-scalar field equations with negative cosmological
constant are found and their local
and global properties are studied. It is found that one of them
represents gravitational collapse where black holes are always formed.
\end{abstract}  

\section{Introduction}

One of the most outstanding problems in gravitation theory is the evolution of
a collapsing massive star, after it has exhausted its nuclear fuel.
However, in order to obtain models for the collapse, we need to 
solve complicated systems of nonlinear differential equations.  
This mathematical complexity is indirectly
revealed by both the relatively few analytic solutions available for study
and the challenges associated with the construction of a good 
numerical code to simulate the collapse process.  Nonetheless, 
work over the last decade has revealed surprisingly rich behaviour in 
even nominally simple ({\it e.g.} spherically symmetric) 
systems\cite{Chop93}.  These 
efforts have led to important advances in our understanding of the 
process of black hole formation and the presence of ``critical" behaviour
in these dynamical, gravitating systems.  Indeed, this behaviour has been
shown to be present in a great variety of systems\cite{Gun00}.

The studies of non-linearity of the Einstein field equations near
the threshold of
black hole formation reveal very rich phenomena \cite{Chop93},
which are quite similar to critical phenomena in Statistical
Mechanics and Quantum Field Theory \cite{Golden}. In particular,
by numerically studying the gravitational collapse of a massless
scalar field in $3+1$-dimensional spherically symmetric
spacetimes, Choptuik found that the mass of such formed black
holes takes a scaling form,
  \bq
  \lb{1.1}
  M_{BH} = C(p)\left(p -p^{*}\right)^{\gamma},
  \eq
where $C(p)$ is a constant and depends on the initial data, and
$p$ parameterises a family of initial data in such a way that when
$p > p^{*}$  black holes are formed, and when $p < p^{*}$ no black
holes are formed. It was shown that, in contrast to $C(p)$, the
exponent $\gamma$   is universal to all the families of initial
data studied. Numerically it was determined as $\gamma \sim 0.37$.
The solution with $p = p^{*}$, usually called the critical
solution, is found also universal. Moreover, for the massless
scalar field it is periodic, too. Universality  of the critical
solution and exponent, as well as the power-law scaling of the
black hole mass all have given rise to the name {\em Critical
Phenomena in Gravitational Collapse}. Choptuik's studies were soon
generalised to other matter fields \cite{Gun00,Wang01}, and now
the following seems clear: (a) There are two types of critical
collapse, depending on whether the black hole mass takes the
scaling form (\ref{1.1}) or not. When it takes the scaling form,
the corresponding collapse is called Type $II$ collapse, and when
it does not it is called Type $I$ collapse. In the type $II$
collapse, all the critical solutions found so far have either
discrete self-similarity (DSS) or homothetic self-similarity
(HSS), depending on the matter fields. In the type $I$ collapse,
the critical solutions have neither DSS nor HSS. For certain
matter fields, these two types of collapse can co-exist. (b) For
Type $II$ collapse, the corresponding exponent is universal only
with respect to certain matter fields. Usually, different matter
fields have different critical solutions and, in the sequel,
different exponents. But for a given matter field the critical
solution and the exponent are universal. So far, the
studies have been mainly restricted to spherically symmetric case
and their non-spherical linear perturbations. Therefore, it is not
really clear whether or not the critical solution and exponent are
universal with respect to different symmetries of the
spacetimes \cite{Cho03,Wang03}. (c)  A critical solution for both of the two
types has one and only one unstable mode. This now is considered as one
of the main criteria for a solution to be critical. (d) The
universality of the exponent is closely  related to the last
property. In fact, using dimensional analysis \cite{Even} one can
show that
  \bq
  \lb{1.2}
  \gamma = \frac{1}{\left|k_{1}\right|},
  \eq
where $k_{1}$ denotes the unstable mode.

Some more recent work has been taken the  
consideration of the gravitational collapse of a minimally coupled 
scalar field in the presence of a cosmological constant but in a lower 
dimensional spacetime, namely $2+1$.  There are several motivations for 
studying such a  
model beyond the intrinsic interest in examining critical behaviour in another
system.  Among these is the recent flurry of work on anti-de Sitter (AdS) 
spacetimes stemming from the AdS/CFT conjecture.  This conjecture assumes
a correspondence between 
the gravitational physics in an AdS spacetime and the physics of
a conformal field theory on the boundary of AdS.  Hence, 
understanding AdS spacetimes can potentially 
yield insight into Super-Yang-Mills theory (and vice versa).  Another 
motivation for studying $2+1$ scalar field collapse is partly the relative 
simplification that results in going from $3+1$ to $2+1$ dimensional 
gravity.  By itself, this would not necessarily be that compelling, but there 
are, of course, some intriguing solutions in $2+1$ such as the BTZ black 
hole \cite{btz} that closely parallel the black hole solutions of $3+1$ gravity.  
Earlier work has considered the question of gravitational collapse to a 
BTZ black hole, but using either null fluid or dust as the collapsing 
matter \cite{H94,PS95}.

Lately, Pretorius and Choptuik (PC) \cite{PC00} studied
gravitational collapse of a massless scalar field in an anti-de
Sitter background in $2+1$-dimensional spacetimes with circular
symmetry, and found that the  collapse exhibits  critical
phenomena and the mass of such formed black holes takes the
scaling form of equation (\ref{1.2}) with $\gamma = 1.2 \pm 0.02$, which
is different from that of the corresponding $3+1$-dimensional
case.  In addition, the critical solution is also different, and,
instead of having DSS, now has HSS. The above results were
confirmed by independent numerical studies 
of Husain and Olivier \cite{HO01}. However,
their exponent,  $\gamma \sim
0.81$,  is quite different from the one obtained by  PC. It is not
clear whether the difference is due to numerical errors or to some
unknown physics.

In 2+1 dimensional spacetimes, the cosmological 
constant plays a fundamental role in black holes
structure. The BTZ solution itself needs a negative
cosmological constant and 
Ida \cite{semvolta} demonstrated 
that, in fact, black holes in this class of spacetimes 
are formed only in the presence of negative cosmological constant. 
To contribute to the above open problems, in 
this paper we shall present the general geodesic
solution of the Einstein-massless scalar field equations in $2+1$
dimensional circularly symmetric spacetimes with negative cosmological
constant, and then study their local and global properties.
The study of their linear perturbations will be considered somewhere else.

\section{The Field Equations}
 
The Einstein's equation for a massless scalar field with
cosmological constant can be written as
\bq
\lb{Rmunu}
R_{\mu \nu} = \phi_{,\mu} \phi_{,\nu} - 2 g_{\mu \nu} \Lambda,
\eq
\noindent where the comma denotes partial differentiation.

The general metric for circularly symmetric spacetimes is given by
\bq
\lb{ds2}
ds^{2} = A^2(r,t)dt^2-B^2(r,t)dr^2-C^2(r,t)d\theta^2.
\eq

Thus, the non-null components of the Ricci tensor are
\bqn
\lb{R_0_0}
R_{00}& = &\frac{A^2}{B^2} \left( \frac{A'}{A} \frac{B'}{B} - \frac{A''}{A} -
\frac{A'}{A} \frac{C'}{C} \right) +
\frac{\ddot B}{B} - \frac{\dot A}{A} \frac{\dot B}{B} + \frac{\ddot C}{C}
-\frac{\dot A}{A} \frac{\dot C}{C},
\\
\lb{R_1_0}
R_{10} & = &\frac{\dot C'}{C} - \frac{\dot B}{B} \frac{C'}{C} -
\frac{\dot C}{C} \frac{A'}{A},
\\
\lb{R_1_1}
R_{11}& = &\frac{C''}{C} + \frac{A''}{A} - \frac{A'}{A} \frac{B'}{B} -
\frac{B'}{B} \frac{C'}{C} -
\frac{B^2}{A^2} \left( \frac{\ddot B}{B} - \frac{\dot A}{A} \frac{\dot
B}{B} +
\frac{\dot B}{B} \frac{\dot C}{C} \right),
\\
\lb{R_2_2}
R_{22}& = &\frac{C^2}{B^2} \left( \frac{C''}{C} + \frac{A'}{A} \frac{C'}{C} -
\frac{B'}{B} \frac{C'}{C} \right)-
\frac{C^2}{A^2} \left( \frac{\ddot C}{C} - \frac{\dot A}{A} \frac{\dot
B}{B} +
\frac{\dot B}{B} \frac{\dot C}{C} \right).
\eqn
In the following we will study the geodesic solutions.
In this case,  these are given in a referential where
the fluid has null acceleration.
This condition implies that
\bq
\lb{geoc}
A=1.
\eq
The non-null components of the Ricci tensor 
for the metric (\ref{ds2}) with the condition
(\ref{geoc}) are  
\bqn
\lb{R_0^0a}
R_0^0 & = & \frac{\ddot B}{B} + \frac{\ddot C}{C},
\\
\lb{R_1^0a}
R_1^0& = &\frac{\dot C'}{C} - \frac{\dot B}{B} \frac{C'}{C},
\\
\lb{R_1^1a}
R_1^1 & = & \frac{\ddot B}{B} + \frac{\dot B}{B} \frac{\dot C}{C}-
\frac{1}{B^2} \left( \frac{C''}{C} - \frac{B'}{B} \frac{C'}{C} \right),
\\
\lb{R_2^2a}
R_2^2 & = &\frac{\ddot C}{C} + \frac{\dot B}{B} \frac{\dot C}{C}-
\frac{1}{B^2} \left( \frac{C''}{C} - \frac{B'}{B} \frac{C'}{C} \right).
\eqn
Above, the prime means partial differentiation in relation to the radial
coordinate, and the 
symbol dot means time differentiation.

Considering  now equation (\ref{Rmunu}), 
the explicit form of the field equations is
\bqn
\lb{R_0^0} &&
R_0^0 + 2\Lambda  =  - \dot \phi^2,
\\
\lb{R_1^1+R_2^2} & &
R_1^1 + R_2^2 + 4\Lambda  =  0,
\\
\lb{R_1^0} & &
R_1^0  =  0,
\\
\lb{R_1^1-R_2^2} & &
R_1^1 - R_2^2 =  0.
\eqn
From (\ref{R_0^0a}) and (\ref{R_0^0}) we have
\bq
\lb{phi0}
\dot \phi^2 = \frac{\ddot B}{B} + \frac{\ddot C}{C} + 2\Lambda.
\eq
On the other hand, equations (\ref{R_1^1a})-(\ref{R_2^2a}) and 
(\ref{R_1^1-R_2^2}) furnish
\bq
\lb{B..-C..}
\frac{\ddot B}{B} - \frac{\ddot C}{C} = 0.
\eq
Equations (\ref{R_1^1a})-(\ref{R_2^2a}) and (\ref{R_1^1+R_2^2}) yield
\bq
\lb{B..+C..}
-\frac{2}{B^2} \left( \frac{C''}{C} -
\frac{B'}{B} \frac{C'}{C} \right)+
\left( \frac{\ddot C}{C} + \frac{\ddot B}{B}+
2 \frac{\dot B}{B} \frac{\dot C}{C} \right)
+ 4\Lambda = 0.
\eq
Finally equations (\ref{R_1^0a}) and (\ref{R_1^0}) give
\bq
\lb{R_1^0b}
\frac{C'}{C} \left( \frac{\dot C'}{C'} - \frac{\dot B}{B} \right) = 0.
\eq
The above equation 
is satisfied for two different relations for the function $C$
\bqn
 (i)  &  \; \; \; \; \;&   \lb{C'C} 
{C'} = 0, \\
(ii) & \; \; \; \; \; &
\lb{C.'C'}
\frac{\dot C'}{C'} = \frac{\dot B}{B} \frac{\dot C}{C}.
\eqn
Henceforth, we will study them separately.

\section{Solutions for the Field Equations}

As we will see below, equation (\ref{C'C}) yields
only one family of solutions, while equation (\ref{C.'C'})
ramifies into two branches.

\bigskip

\noi {\bf Case i) Solutions for C'=0}

\medskip

\noi From (\ref{C'C}) we readily have 
\bq
\lb{Ct}
C=C(t).
\eq
Substituting (\ref{C'C}) into (\ref{B..+C..}) we obtain
\bq
\lb{B..+C..a}
\frac{\ddot B}{B} +
\left( \frac{\ddot C}{C} + 2
\frac{\dot B}{B} \frac{\dot C}{C} \right)
+ 4\Lambda = 0,
\eq
that can be rewritten as
\bq
\lb{B..+C..b}
(B C)^{..} + 4\Lambda B C = 0,
\eq
which furnishes
\bq
\lb{BCa}
B C = c_1(r) e^{2\sqrt{-\Lambda}t} + c_2(r) e^{-2\sqrt{-\Lambda}t}.
\eq
On the other hand,
equations (\ref{B..+C..}) and (\ref{B..-C..}) 
leads to
\bq
\lb{C..+B.C.}
\frac{\ddot C}{C} + \frac{\dot B}{B} \frac{\dot C}{C} =
- 2\Lambda,
\eq
From equation (\ref{C'C}) we can also readily write
\bqn
\lb{cf1}
\frac{\dot C}{C} & = & f_1(t),
\\ \lb{cf2}
\frac{\ddot C}{C} & = & f_2(t).
\eqn
Setting the above definitions into equation (\ref{C..+B.C.}) we have
\bq
\frac{\dot B}{B} = -\frac{2\Lambda + f_2(t)}{f_1(t)},
\eq
\noi where $f_2(t)$ $=$ ${\dot f}_1(t)$ $+$ ${f^2_1}(t)$,
whose solution is
\bq
\lb{Bf}
 B =e^{ -2\Lambda \int \frac{dt}{f_1(t)} - \int f_1 dt  - \ln f_1(t) + g(r)} ,
\eq
where $g(r)$ is an arbitrary function of $r$.
Moreover, the integration of equation (\ref{cf1}) furnishes
\bq
\lb{Cf}
 C = e^{\int f_1(t) dt} .
\eq
The relation (\ref{R_0^0}) for the scalar field is now
\bq
\dot \phi^2 = 2[ \dot f_1(t) + f_1(t)^2 + \Lambda ].
\eq
Considering equations (\ref{Bf})-(\ref{Cf}) into
equation (\ref{B..+C..a}) we can show that $f_1(t)$ must obey
\bq
\lb{f1..}
4\Lambda^2 + 4\Lambda f_1(t)^2 + 6\Lambda \dot f_1(t) - f_1(t) \ddot f_1(t)
+ 2\dot f_1(t)^2 = 0.
\eq
Below we will use it as an integration condition.

From equations (\ref{BCa}), (\ref{Bf}) and (\ref{Cf}) we must
have $c_1(r) = a c_2(r)$, where $a$ is a constant. Then
\bq
\lb{f1.}
\dot f_1 + \frac{2 \sqrt{-\Lambda}
\left( e^{2\sqrt{-\Lambda}t} - a e^{-2\sqrt{-\Lambda}t} \right)}
{e^{2\sqrt{-\Lambda}t} + a e^{-2\sqrt{-\Lambda}t} } f_1 + 2\Lambda = 0,
\eq
which can be solved giving us
\bq
\lb{f1}
f_1(t)=\frac{ a \Lambda + b \sqrt{-\Lambda} e^{2\sqrt{-\Lambda}t} -
\Lambda e^{4t\sqrt{-\Lambda}} }
{ \sqrt{-\Lambda} ( e^{4t\sqrt{-\Lambda}} + a)},
\eq
where $b$ is an integration constant. 
This solution satisfies readily the condition
(\ref{f1..}). Then, the resulting spacetime can be written as
\bqn
\lb{mc'}
ds^2 & = & dt^2 
-\left(e^{2\sqrt{-\Lambda} t} + ae^{-2\sqrt{-\Lambda} t}\right)
\\ && \nb  \; \times
\left[-\Lambda e^{-\frac{b}{2\sqrt{-\Lambda a}}
\arctan \left( \frac{1}{\sqrt{a}}e^{2\sqrt{-\Lambda}t} \right)} dr^2
+ e^{\frac{b}{2\sqrt{-\Lambda a}}
\arctan \left( \frac{1}{\sqrt{a}}e^{2\sqrt{-\Lambda}t} \right)} d\theta^2
\right] ,
\eqn
with $a>0$. The associated scalar field is given by
\bq
\lb{phi}
\phi(t)=\pm \frac{\sqrt{2}}{4}
\sqrt{ -\frac{16a\Lambda - b^2}{a\Lambda}} \arctan \left(
{{\sqrt{a}e^{2\sqrt{-\Lambda}t}}}\right) + \phi_0,
\eq
\noi Thus, we have an imaginary scalar field  and besides we 
can easily show that we do not have singularity formation.

\bigskip

\noi {\bf Case ii) Solutions for $\bf C' \ne 0$}

\medskip

\noi From equation (\ref{C.'C'}) we have
\bq
\lb{CB}
\left( \frac{C'}{B} \right)^{.} = 0,
\eq
that can be readily solved giving
\bq
\lb{C'B}
C' = B h(r).
\eq
Differentiating twice in relation to $t$ we get
\bq
\frac{\ddot B}{B} = \frac{\ddot C'}{C'}.
\eq
Taking this last relation into
equation (\ref{B..-C..}) we find
\bq
\lb{53}
\frac{\ddot C}{C} - \frac{\ddot C'}{C'} = 0,
\eq
which can be rewritten as
\bq
\frac{\ddot C}{C'} \left( \frac{C'}{C} - \frac{\ddot C'}{\ddot C} \right)= 0.
\eq
Thus, we have two possibilities:
\bqn
ii.a)  \; \; & &\lb{C..C} \; \; 
\frac{\ddot C}{C} = 0 \\
ii.b)  \; \; & & \; \; \lb{C..'C..}
\frac{ C'}{C} - \frac{\ddot C'}{\ddot C} =0.
\eqn
Below we will analyse each of them separately. 

\bigskip
 
\noi {\bf  Case ii.a)  Solutions for  $\bf C' \ne 0$ and $\bf \ddot C = 0$}

\medskip 

From equation (\ref{C..C}) we have
\bq
\lb{Chr}
C= h_1(r) t + h_2(r).
\eq
 
Using equation (\ref{C'B}) we get
\bq
\lb{Bhr}
B = \frac{h_1' t + h_2'}{h}.
\eq
Substituting  equations (\ref{Bhr})-(\ref{Chr}) into (\ref{B..+C..}) we get
\bq
h_1'h_1-hh'+2\Lambda(h_1'h_1t^2+h_1'h_2t+h_2'h_1t+h_2'h_2)=0,
\eq
That yields the following system of equations
\bq
h_1'h_1-hh'+2\Lambda h_2'h_2=0,
\eq
\bq
2\Lambda h_1'h_1=0,
\eq
\bq
2\Lambda(h_1'h_2+h_2'h_1)=0.
\eq
This system presents two solutions that are $h_1$ $=$ $0$ and
$h_1'$ $=$  $0$.

\medskip

\noi {\bf Case ii.a.1 Solution for $ h_1=0 $}

\smallskip

\noi This implies $ h $ 
$=$ $\pm$  $\sqrt{2\Lambda h_2 + h_0}$ which
furnishes $ \dot \phi$ $\pm$ $\sqrt{2\Lambda}$ and thus
\bqn
C&=&h_2(r), \nonumber \\
B&=&\frac{h_2'(r)}{g(r)}, \nonumber \\
\phi&=&\pm t \sqrt{2\Lambda} + \phi_0, \nonumber \\
R&=&-4\Lambda,
\eqn
\noi where $R$ is the Ricci scalar. We have then
a static metric and an imaginary scalar field.

\medskip

\noi {\bf Case ii.a.2) Solution for $h_1'=0 $} 

\smallskip

\noi This implies $ h_2$ $=$ constant and thus  $B=0 $, 
which is also unacceptable for obvious reasons.

\bigskip

\noi {\bf Case ii.b) Solutions for  $\bf C' \ne 0$ and $\bf \ddot C \ne 0$}

\medskip

Substituting equation (\ref{C'B}) into equation 
(\ref{B..-C..}) we can write
\bq
\lb{C..C..}
\frac{\ddot C'}{\ddot C}  =  \frac{C'}{C}, \eq
then
\bq
\lb{C..hC}
\ddot C =  q(t) C.
\eq
Considering equations (\ref{C'B})-(\ref{B..-C..}) into
equation (\ref{R_1^1+R_2^2}) we have 
\bq
\lb{qCh}
2\left[q(t) +2\Lambda\right]C'C + 2 \dot C' \dot C  =  2 h h',\eq
whose integration in $r$ furnishes 
\bq
\lb{qCha}
\left[q(t) +2\Lambda \right] C^2 + \dot C^2 - h^2 = q_1(t) .
\eq
Now taking equation (\ref{C'B}) into equation (\ref{phi0}), we obtain
\bq
\lb{C.2}
{\dot \phi}^2   = \frac{\ddot C'}{C'} + \frac{\ddot C}{C} + 2\Lambda,
\eq
and, with equation (\ref{C..hC}), we can write
\bq
\dot \phi^2  =  2[q(t) + \Lambda],
\eq
Putting $q(t)$ from equation (\ref{C..hC}) into equation
(\ref{qCha}) we have now
\bq
\lb{nao}
(C^2)^{..} + 4\Lambda C^2 = 2h^2 + 2q_1,
\eq
\noi the integration of the above equation holds
\bqn
\lb{C2}
C^2  & = & e^{2\sqrt{-\Lambda}t} \left[ \frac{1}{2\sqrt{-\Lambda}}
\int q_1(t)e^{-2\sqrt{-\Lambda}t} dt + h_1(r) \right] \nb\\
&  & \;  -e^{-2\sqrt{-\Lambda}t} \left[ \frac{1}{2\sqrt{-\Lambda}}
\int q_1(t)e^{2\sqrt{-\Lambda}t} dt + h_2(r) \right] +
\frac{h(r)^2}{2\Lambda}.
\eqn
Finally, substituting  equation (\ref{C2}) into
equation (\ref{C..hC}) we can find
\bqn
\lb{q}
q&=& \frac{ 2(-\Lambda) \left[
e^{2\sqrt{-\Lambda}t} \alpha(t,r) -
e^{-2\sqrt{-\Lambda}t} \beta(t,r)  \right] + q_1(t)}
{ e^{2\sqrt{-\Lambda}t} \alpha(t,r) -
e^{-2\sqrt{-\Lambda}t} \beta(t,r)  +
\frac{h(r)^2}{2\Lambda}} \nonumber \\
& &- \frac{ (-\Lambda) \left[
e^{2\sqrt{-\Lambda}t} \alpha(t,r) +
e^{-2\sqrt{-\Lambda}t} \beta(t,r) \right]^2 }
{ \left[ e^{2\sqrt{-\Lambda}t} \alpha(t,r) -
e^{-2\sqrt{-\Lambda}t} \beta(t,r) +
\frac{h(r)^2}{2\Lambda} \right]^2 },
\eqn
where
\bqn
\lb{alpha}
\alpha (t,r) & = & \frac{1}{2\sqrt{-\Lambda}}
F_1(t) + h_1(r), \\
\lb{beta}
\beta (t,r) & = & \frac{1}{2\sqrt{-\Lambda}}
F_2(t) + h_2(r), 
\\
F_1(t) &= &\int q_1(t)e^{-2\sqrt{-\Lambda}t} dt, 
\eqn
and
\bq
F_2(t) = \int q_1(t)e^{2\sqrt{-\Lambda}t} dt .
\eq
Since $h_1 $ $=$ $h_2$ $=$ $h$ $=$ constant, we have $C'$ $=$ $0$, the unique
solution for this case is

\bq
d_0 h^2=h_1 d_1 = h_2 d_2 \;\; {\rm{and}} \;\; q_1(t)=0,
\eq
which give us the general solution for the Case ii,
\bqn
C^2 & = & h(r)^2 \left( \frac{d_0}{2\Lambda} + d_1e^{2\sqrt{-\Lambda}t}+
d_2e^{-2\sqrt{-\Lambda}t} \right),
\\
B^2 & = & \frac{h'(r)^2}{\sqrt{d_0 h(r)^2+d_3}}
 \left( \frac{d_0}{2\Lambda} + d_1e^{2\sqrt{-\Lambda}t}+
d_2e^{-2\sqrt{-\Lambda}t} \right),
\\
\lb{phig}
\phi& = &-\frac{1}{\sqrt{-\Lambda}} \arctan \left(
\frac{4d_1 e^{2\sqrt{-\Lambda}t} + d_0}
{\sqrt{16d_1d_2\Lambda^2 - d_0^2}} \right) + \phi_0,
\eqn
where $d_0$, $d_1$, $d_2$, $d_3$ and $\phi_0$ are constants.

Doing a coordinate transformation, such as
\bq
\bar r = \frac{1}{\sqrt{d_0}} \ln \left( \sqrt{d_0} \; h(r) + \sqrt{d_0 h(r)^2 + d_3} \right) + {\rm{constant}}
\eq
we can rewrite the metric, dropping the bar notation for the new
coordinate, as
\bqn
\lb{ds2final}
ds^2 & = & dt^2- \left( \frac{d_0}{2\Lambda} + d_1e^{2\sqrt{-\Lambda}\;t}+
d_2e^{-2\sqrt{-\Lambda}\;t} \right) 
\\ \nonumber && \; \times
\left[ dr^2 + \frac{1}{4 d_0} ( e^{\sqrt{d_0}\; r}-
d_3 e^{-\sqrt{d_0}\; r} )^2 d\theta^2 \right].
\eqn
This is then the unique physical solution. Below we will study it in  more 
details.

\section{Study of the Global Structure of the Physical Solution}

For gravitational collapse, we impose the following conditions at $r=0$:
\begin{description}

\item ($i$) There must exist a symmetry axis, which can be expressed as
\bq
\lb{cd1}
X \equiv \left|\xi^{\mu}_{(\theta)}\xi^{\nu}_{(\theta)}
g_{\mu\nu} \right| \rightarrow 0,
\eq
as $r \rightarrow 0$, we have chosen the radial coordinate
such that the axis is located at $r = 0$, and
$\xi^{\mu}_{(\theta)}$ is the Killing vector with a close orbit,
and given by $\xi^{\alpha}_{(\theta)}\partial_{\alpha} =
\partial_{\theta}$.
 
\item ($ii$) The spacetime near the symmetry axis is locally flat, which
can be written as \cite{Kramer80}
\bq
\lb{cd2}
\frac{X_{,\alpha}X_{,\beta} g^{\alpha\beta}}{4X}
\rightarrow - 1,
\eq
as $r \rightarrow 0$.  Note that solutions failing to satisfy this
condition sometimes are also acceptable. For example, when the
left-hand side of the above equation approaches a finite constant,
the singularity at $r = 0$ may be related to a point-like particle \cite{VS}. 
 
\item ($iii$) No closed timelike curves (CTC's). In spacetimes with
circular symmetry, CTC's can be easily introduced. To ensure
their  absence, we assume that the condition
\bq
\lb{cd3}
\xi^{\mu}_{(\theta)}\xi^{\nu}_{(\theta)}g_{\mu\nu} < 0,
\eq
holds in the whole spacetime.

\end{description} 

Applying these conditions we can easily see from the metric 
(\ref{ds2final}) that 
the regularity conditions on the axis $r = 0$ requires  $d_{3} = 1$. 

The geometrical radius is given by
\bq
R_g \equiv \sqrt{g_{\theta\theta}}
 = \frac{1}{2 \sqrt{d_0}}\sqrt{\frac{d_0}{2\Lambda}+d_1e^{2\sqrt{-\Lambda}t}+
d_2e^{-2\sqrt{-\Lambda}t} } \left( e^{r\sqrt{d_0}}- e^{-r\sqrt{d_0}} \right).
\eq
In order to have a real positive geometrical radius we may have two 
possibilities:
\begin{description}
\item a) $d_0>0 ,\;\; \frac{d_0}{2\Lambda}+d_1e^{2\sqrt{-\Lambda}t}+
d_2e^{-2\sqrt{-\Lambda}t} >0 ,\;\; d_3 \le 1$,
\item b) $d_0<0 ,\;\; \frac{d_0}{2\Lambda}+d_1e^{2\sqrt{-\Lambda}t}+
d_2e^{-2\sqrt{-\Lambda}t} <0 ,\;\; d_3=-1$,
\end{description}
but this last condition violates the regularity condition.

The condition for a collapse is given by
\bq
\dot R_g= \frac{\sqrt{-\Lambda}}{4\sqrt{d_0}} 
\frac{d_1 e^{2\sqrt{-\Lambda}t} - d_2 e^{-2\sqrt{-\Lambda}t}}
{\sqrt{ \frac{d_0}{2\Lambda} +d_1 e^{2\sqrt{-\Lambda}t} + d_2 e^{-2\sqrt{-\Lambda}t}} } ( e^{r\sqrt{d_0}}- e^{-r\sqrt{d_0}} ) < 0.
\eq
Thus, we have 
\bq
d_0>0 ,\;\; d_1<0 ,\;\; d_2>0,
\eq
without any restriction in the time coordinate. 

The condition to have a 
real geometrical radius gives an additional constrain
written as
\begin{description}
\item a) for $d_0>0, \;\; d_2 > 0$ and $d_1 < 0$, 
 $ \frac{d_0}{2 \Lambda} - |d_1| + |d_2| \ge 0 , $ 
\item b) for $d_0<0, \;\; d_2 < 0$ and $d_1 > 0$,
 $-\frac{ |d_0 |}{2 \Lambda} + |d_1| - |d_2| \ge 0. $
\end{description}

\noi The second conditions represent an expansion instead of a collapse, so
we consider hereinafter only the first ones.

The energy density of the fluid is given by
\bq
\lb{rho}
\rho=\frac{ -\Lambda \left[ \left( \frac{d_0}{2\Lambda} +d_1 e^{2\sqrt{-\Lambda}t} + d_2 e^{-2\sqrt{-\Lambda}t} \right)^2 + \frac{d_0^2}{4\Lambda^2} - 4 d_1 d_2 \right] } {\left( \frac{d_0}{2\Lambda} +d_1 e^{2\sqrt{-\Lambda}t} + d_2 e^{-2\sqrt{-\Lambda}t} \right)^2},
\eq
and it can be shown that $\rho$ is positive for $\Lambda<0$, $d_1<0$ 
and $d_2>0$, and whose denominator can vanish at 
\bq
t_{sing}=\frac{1}{2\sqrt{-\Lambda}} \ln \left(
\frac{-d_0 \pm \sqrt{d_0^2-16d_1d_2 \Lambda^2}}{4d_1 \Lambda} \right),
\eq
which represents the instant of the singularity formation.
Here the positive sign is for $d_0 >0$ and the negative sign
is for $d_0<0$. 

It is easily shown that 
$R_g$ vanishes for the singularity, i.e., $R_g(t=t_{sing}) = 0$. 

The pressure is given by
\bq
\lb{p}
p=\frac{ -\Lambda \left[ -3\left( \frac{d_0}{2\Lambda} +d_1 e^{2\sqrt{-\Lambda}t} + d_2 e^{-2\sqrt{-\Lambda}t} \right)^2 + \frac{d_0^2}{4\Lambda^2} - 4 d_1 d_2 \right] } {\left( \frac{d_0}{2\Lambda} +d_1 e^{2\sqrt{-\Lambda}t} + d_2 e^{-2\sqrt{-\Lambda}t} \right)^2}.
\eq

The expansion of the congruence of null outgoing and 
ingoing geodesics can be written, respectively, as 
\bqn
\lb{theta_l}
\theta_l  & \equiv &
\frac{F}{R_g} \left( \frac{1}{A} {R_g},_t + \frac{1}{B} {R_g},_r
 \right)
\\
&= & \nonumber
\frac{\sqrt{-\Lambda}} {\sqrt{\frac{2d_0}{\Lambda}+d_1e^{2\sqrt{-\Lambda}t}+d_2e^{-2\sqrt{-\Lambda}t}}} \left( d_1 e^{2\sqrt{-\Lambda}t}-d_2 e^{-2\sqrt{-\Lambda}t} \right)
\\ & & \nonumber +\frac{1}{2} \left( e^{r\sqrt{d_0}}+ e^{-r\sqrt{d_0}} \right),
\\
\lb{theta_n}
\theta_n 
 & \equiv &
\frac{G}{R_g} \left( \frac{1}{A} {R_g},_t - \frac{1}{B} {R_g},_r
 \right)\\ \nonumber
& = & 
\frac{\sqrt{-\Lambda}} {\sqrt{\frac{2d_0}{\Lambda}+d_1e^{2\sqrt{-\Lambda}t}+d_2e^{-2\sqrt{-\Lambda}t}}} \left( d_1 e^{2\sqrt{-\Lambda}t}-d_2 e^{-2\sqrt{-\Lambda}t} \right)  \\ && \nonumber
-\frac{1}{2} 
\left( e^{r\sqrt{d_0}}  + e^{-r\sqrt{d_0}} \right).
\eqn
where $F$ and $G$ are always positive \cite{CFJW04}.
The apparent horizon $r_{AH}$ is located at the hypersurface
$\theta_n\theta_l=0$ \cite{Pen68}. Besides,
equation (\ref{theta_n}) shows us  that $\theta_n$ is always negative,
as it would be expected in a collapse process. It can be also shown that
for $r$ $>$ $r_{AH}$, we have $\theta_l$ positive, while in the region where
 $r$ $<$ $r_{AH}$,  the expansion $\theta_l$ is always negative. 
This means that the collapse always forms black holes.

Summarizing, the solution obtained here satisfies the regularity conditions
at the origin and presents a real and positive geometric radius, which
allways decreases with the time. A singularity is formed in a finite time, 
where the geometric radius vanishes.

\section{Conclusion}

In this paper, we found all the geodesic solutions of the
Einstein-massless-scalar field equations with negative cosmological constant
in the $(2+1)$-dimensional
spacetimes with circular symmetry. From these,
we have shown that only one of the
solutions, given by the equations (\ref{phig}) and (\ref{ds2final}),
can represent gravitational collapse of the scalar field.
The collapse always forms black holes. We intend to analyse
the perturbations of this solution in order to investigate the possibility of a critical type I collapse and present the results as soon as possible.

\section*{ Acknowledgements}

The financial assistance from UERJ (JFVdaR), and FAPERJ/UERJ (MFAdaS) is
gratefully acknowledged. The author (RC) also
gratefully acknowledges the financial 
support from FAPERJ (no. E-26/171.754/2000 and E-26/171.533/2002).
We would also like to thank the useful 
and helpful discussions with Dr. Anzhong Wang.

\end{document}